\documentclass[10pt]{llncs}
\title{Security Type Systems as Recursive Predicates\thanks{This work was supported by the DFG project Ni 491/13--2, 
part of the DFG priority program Reliably Secure Software Systems (RS3).}}
\author{Andrei Popescu}
\institute{Technische Universit\"{a}t M\"{u}nchen 
           }

\usepackage{graphicx}
\usepackage{alltt}
\usepackage{url}
\usepackage{mathptmx}
\usepackage{amssymb}

\newcommand{\comment}[1]{}

\newcommand{\ov}{\overline}

\renewcommand{\phi}{\varphi}

\renewcommand{\partial}{\rightharpoondown}

\newcommand{\ra}{\rightarrow}

\newcommand{\LRA}{\Longrightarrow}

\newcommand{\inferN}[3]{\mbox{$\displaystyle \frac{#2}{#3}$}\mbox{\rm\textsf{(#1)}}} 

\newcommand{\Vars}{\mbox{\rm\textsf{\small Vars}}}
\newcommand{\low}{\mbox{\rm\textsf{\small low}}}
\newcommand{\wlow}{\mbox{\rm\textsf{\small wlow}}}
\newcommand{\high}{\mbox{\rm\textsf{\small high}}}
\newcommand{\fhigh}{\mbox{\rm\textsf{\small fhigh}}}

\newcommand{\maxTp}{\mbox{\rm\textsf{\small max\hspace*{-0.04ex}Tp}}}
\newcommand{\maxWtp}{\mbox{\rm\textsf{\small max\hspace*{-0.04ex}Wtp}}}
\newcommand{\minTp}{\mbox{\rm\textsf{\small min\hspace*{-0.04ex}Tp}}}
\newcommand{\minRtp}{\mbox{\rm\textsf{\small min\hspace*{-0.04ex}Rtp}}}
\newcommand{\minTRtp}{\mbox{\rm\textsf{\small min\hspace*{-0.04ex}T\hspace*{-0.04ex}Rtp}}}

\newcommand{\noWhile}{\mbox{\rm\textsf{\small no\hspace*{-0.04ex}While}}}

\newcommand{\mayT}{\mbox{\rm\textsf{\small may\hspace*{-0.04ex}T}}}

\newcommand{\safe}{\mbox{\rm\textsf{\small safe}}}

\newcommand{\pres}{\mbox{\rm\textsf{\small pres}}}

\newcommand{\cpt}{\mbox{\rm\textsf{\small cpt}}}

\newcommand{\hi}{\mbox{\rm\textsf{\small hi}}}

\newcommand{\lo}{\mbox{\rm\textsf{\small lo}}}

\newcommand{\secc}{\mbox{\rm\textsf{\small sec}}}

\newcommand{\siso}{\mbox{\rm\textsf{\small siso}}}

\newcommand{\True}{\mbox{\rm\textsf{\small True}}}

\newcommand{\False}{\mbox{\rm\textsf{\small False}}}

\newcommand{\discr}{\mbox{\rm\textsf{\small discr}}}

\newcommand{\Seq}{\mbox{\rm\textsf{\small Seq}}}

\newcommand{\If}{\mbox{\rm\textsf{\small If}}}

\newcommand{\Par}{\mbox{\rm\textsf{\small Par}}}

\newcommand{\While}{\mbox{\rm\textsf{\small While}}}

\newcommand{\defn}{\mbox{$\;\,\equiv\;\,$}}

\newcommand{\zobis}{\mbox{$\;\approx_{\tiny \textsf{01}}\,$}}

\newcommand{\wbis}{\mbox{$\;\approx_{\tiny \textsf{W}}\,$}}
\newcommand{\wbist}{\mbox{$\;\approx_{\tiny \textsf{WT}}\,$}}

\newcommand{\ind}{\mbox{$\;\sim\;$}}

\newcommand{\tst}{\mbox{\textit{tst}}}

\newcommand{\atm}{\mbox{\textit{atm}}}

\newcommand{\com}{\mbox{\bf com}}

\newcommand{\test}{\mbox{\bf test}}

\newcommand{\atom}{\mbox{\bf atom}}

\newcommand{\state}{\mbox{\bf state}}

\newcommand{\intt}{\mbox{\bf int}}
\newcommand{\bool}{\mbox{\bf bool}}

\newcommand{\expp}{\mbox{\bf exp}}

\newcommand{\var}{\mbox{\bf var}}

\include{tikzlib}

\usepackage[T1]{fontenc}
\usepackage[all]{xy}
\usepackage{amsmath}
\usepackage{bussproofs}
\usepackage{cite}
\usepackage{color}
\usepackage[scaled=.9]{helvet}
\usepackage{graphicx}
\usepackage{mathptmx}
\usepackage{url}
\usepackage[scaled=.825]{beramono}

\Roman{section}

\begin{document}
\bibliographystyle{abbrv}




\maketitle

\vspace*{-3.5ex}
\begin{abstract} \rm 
We show how security type systems from the literature of language-based noninterference 
can be represented more directly as predicates defined by structural recursion on the programs.      
In this context, we show how our uniform syntactic criteria from \cite{cpp,jfr} cover 
several previous type-system soundness results.   
\end{abstract}

\section{Security type systems}

 
As in Example 2 from \cite{cpp,jfr},  
we assume 
that atomic statements and tests are built by means of 
expressions applied to variables taken 
from a set $\var$, ranged over by $x,y,z$. Thus, $\expp$, ranged over by $e$, 
is the set of arithmetic expressions (e.g., $x+1$, $x * y + 5$). Then atomic commands $\atm \in \atom$ are assignment 
statements $x := e$ and tests $\tst \in \test$ 
are Boolean expressions built from $\expp$ (e.g., $x > 0$, $x + 1 = y + z$).  
For any expression $e$ and test $\tst$, $\Vars\;e$ and $\Vars\;\tst$  denote their sets of variables.  

States are assignments 
of integers to variables, i.e., the set $\state$ is $\var \ra \intt$.  
Variables are classified as either 
low ($\lo$) or high ($\hi$) by a fixed security level 
function $\secc : \var \ra \{\lo,\hi\}$. We let $L$ be the lattice $\{\lo,\hi\}$, 
where $\lo < \hi$.\footnote{One can also consider the more general 
case of {\em multilevel security}, via 
an unspecified lattice of security levels $L$---however, this brings neither 
much additional difficulty, nor much additional insight, so here 
focus on this 
$2$-level lattice.}  We shall use the standard infima and suprema notations for $L$.     
Then $\!\!\ind\!\!$ is defined as follows: $s \ind t \defn \forall x \in \var.\;\secc\;x = 
\lo \LRA s\;x = t\;x$.

We shall look into type systems from the literature, $::$, assigning security 
levels $l \in \{\lo,\hi\}$, or pairs of security levels, to expressions and commands.
All have in common the following: 
 
Typing of expressions: 
$$e :: \lo \mbox{ \ if \ } \forall x \in \Vars\;e.\;\sec\;x = \lo \hspace*{10ex} e :: \hi \mbox{ \ always}$$ 

Typing of tests (similar): 
$$\tst :: \lo \mbox{ \ if \ } \forall x \in \Vars\;\tst.\;\sec\;x = \lo \hspace*{10ex} \tst :: \hi \mbox{ \ always}$$ 

The various type systems shall differ in the typing of commands.  

But first let us look more closely at their aforementioned common part.  We note that, if an expression or a test has type $l$ and $l \leq k$, 
then it also has type $k$. In other words, the following covariant subtyping rules for tests and expressions hold: 

\begin{center}
$\inferN{SUBTYPE-EXP}
{e :: l\hspace*{3ex} l \leq k}
{e :: k}$
\hspace*{15ex} 
$\inferN{SUBTYPE-TST}
{\tst :: l\hspace*{3ex} l \leq k}
{\tst :: k}$
\end{center}

Thus, the typing of an expression or test is uniquely determined by its {\em minimal type}, 
defined as follows: 
$$\minTp\;e = \bigvee\{\sec\;x.\;x \in \Vars\;e\}
\hspace*{5ex}
\minTp\;\tst = \bigvee\{\sec\;x.\;x \in \Vars\;\tst\}$$

The minimal typing operators can of course recover the original typing relation $::$ as follows: 

\begin{lemma} \rm \label{lem-tp}
The following hold:
\\(1) $e :: l$ iff $\minTp\;e \leq l$.
\\(2) $\tst :: l$ iff $\minTp\;\tst \leq l$.
\end{lemma}

\subsection{Volpano-Smith possibilistic noninterference}

In \cite[\S4]{smiVol-multiThr}, the typing of commands (which we denote by $::_1$) is defined inductively as follows:     
\begin{center}
$\inferN{ASSIGN}
{\sec\;x = l \hspace*{3ex} e :: l}
{(x := e) ::_1 l}$
\hspace*{10ex}
$\inferN{COMPOSE}
{c_1 ::_1 l \hspace*{3ex} c_2 ::_1 l}
{(\Seq\;c_1\;c_2) ::_1 l}$
\\
\vspace*{1.3ex} 
$\inferN{IF}
{\tst ::_1 l \hspace*{3ex} c_1 ::_1 l \hspace*{3ex} c_2 ::_1 l}
{(\If\;\tst\;c_1\;c_2) ::_1 l}$
\hspace*{10ex} 
$\inferN{WHILE}
{\tst ::_1 \lo \hspace*{3ex} c ::_1 l}
{(\While\;\tst\;c) ::_1 \lo}$
\\
\vspace*{1.3ex}
$\inferN{PAR}
{c_1 ::_1 l \hspace*{3ex} c_2 ::_1 l}
{(\Par\;c_1\;c_2) ::_1 l}$
\hspace*{15ex} 
$\inferN{SUBTYPE}
{c ::_1 l\hspace*{3ex} k \leq l}
{c ::_1 k}$
\end{center} 

We think of $c ::_1 l$ as saying: 
\begin{itemize}
\item There is no downwards flow in $c$. 
\item $l$ is a lower bound on the level of the variables that the execution of $c$ writes to. 
\end{itemize}
(This intuition is accurately reflected by Lemma \ref{lem-tp1} below.)

Actually, \cite{smiVol-multiThr} does not explicitly consider a rule like (PAR), and in fact uses parallel composition only at the top 
level.  However, it does require that the thread pool (which can be viewed as consisting of a number of parallel compositions) has 
well-typed threads, which is the same as typing the pool to the minimum of the types of its threads---this is precisely what (PAR) does.  
(Also, in \cite{smiVol-multiThr}, the rule (WHILE) has the assumption $c ::_1 \lo$ rather that $c ::_1 l$---this alternative is of course equivalent,  
thanks to (SUBTYPE).)   

Due to the subtyping rule, here we have a phenomenon dual to the one for expressions and tests: if a command has type $l$ and $k \leq l$, 
then it also has type $k$---thus, the typing of a command, if any, is uniquely determined by its {\em maximal type}.  
The difference from expressions and tests is that such a type may not exist, making it necessary to keep a ``safety" predicate 
during the computation of the maximal type.  
For example, consider the computation of the minimal type of $\If\;\tst\;c_1\;c_2$ according to the (IF) rule:
Assume $l_0$ is the minimal type of $\tst$ and $l_1,l_2$ are the maximal types of $c_1$ and $c_2$, respectively.   
The rule (IF) requires the three types involved in the hypothesis to be equal, and therefore we need to {\em upcast} $l_0$ and {\em downcast} $l_1$ and $l_2$ 
so that we obtain a common type $l$---thus, we need $l_0 \leq l \leq l_1 \wedge l_2$.  Moreover, $l$ has to be as high as possible.  
Such an $l$ of course only exists if $l_0 \leq l_1 \wedge l_2$, and in this case the maximal $l$ is $l_1 \wedge l_2$.  
In summary, the rule (IF) tells us the following: 
\begin{itemize}
\item $\If\;\tst\;c_1\;c_2$ is safe (i.e., type checks) iff $c_1$ and $c_2$ are safe and $l_0 \leq l \leq l_1 \wedge l_2$.
\item If safe, the maximal type of $\If\;\tst\;c_1\;c_2$ is $l_1 \wedge l_2$.  
\end{itemize}

Applying this reasoning to all the rules for $::_1$, we obtain the function $\maxTp_1 : \com \ra L$ 
and the predicate $\safe_1 : \com \ra \bool$ defined recursively on the structure of commands:\footnote{Notice the overloaded, but consistent 
usage of the infimum operator $\wedge$ in both the lattice $L = \{\lo,\hi\}$ and the lattice of truth values $\bool$ (the latter simply meaning the logical ``and").}

\begin{definition} \rm \label{def-safe1}
\begin{itemize}
\item $\safe_1\;(x := e) = (\minTp\;e \leq \sec\;x)$
\item $\maxTp_1\;(x := e) = \sec\;x$
\item $\safe_1\;(\Seq\;c_1\;c_2) = (\safe_1\;c_1 \wedge \safe_1\;c_2)$
\item $\maxTp_1\;(\Seq\;c_1\;c_2) = (\maxTp_1\;c_1 \wedge \maxTp_1\;c_2)$
\item $\safe_1\;(\If\;\tst\;c_1\;c_2) = (\safe_1\;c_1 \wedge \safe_1\;c_2 \wedge (\minTp\;\tst \leq (\maxTp_1\;c_1 \wedge \maxTp_1\;c_2)))$
\item $\maxTp_1\;(\If\;\tst\;c_1\;c_2) = (\maxTp_1\;c_1 \wedge \maxTp_1\;c_2)$
\item $\safe_1\;(\While\;\tst\;c) = (\safe_1\;c \wedge (\minTp\;\tst = \lo))$
\item $\maxTp_1\;(\While\;\tst\;c) = \lo$
\item $\safe_1\;(\Par\;c_1\;c_2) = (\safe_1\;c_1 \wedge \safe_1\;c_2)$
\item $\maxTp_1\;(\Par\;c_1\;c_2) = (\maxTp_1\;c_1 \wedge \maxTp_1\;c_2)$
\end{itemize}
\end{definition}

\begin{lemma} \rm \label{lem-tp1}
The following are equivalent:
\\(1) $c ::_1 l$ 
\\(2) $\safe_1\;c$ and $l \leq \maxTp_1\;c$.
\end{lemma}

{\em Proof idea:}
(1) implies (2): By easy induction on the definition of $::_1$. 
\\(2) implies (1): By easy structural induction on $c$.
\qed

\ \\
Now, let us write:
\begin{itemize}
\item $\low\;e$, for the sentence $\minTp\;e = \lo$ 
\item $\low\;\tst$, for the sentence $\minTp\;\tst = \lo$
\item $\fhigh\;c$ (read ``$c$ finite and high"), for the sentence $\maxTp_1\;c = \hi$
\end{itemize}
(Thus, $\low : \exp \ra \bool$, $\low : \test \ra \bool$ and $\fhigh : \com \ra \bool$.)

Then, immediately from the definitions of $\minTp$ and $\maxTp_1$ (taking advantage of the fact that $L = \{\hi,\lo\}$) we have the following:
\begin{itemize}
\item $\low\;e = (\forall x \in \Vars\;e.\sec\;x = \lo)$
\item $\low\;\tst = (\forall x \in \Vars\;\tst.\sec\;x = \lo)$
\end{itemize}

\begin{itemize}
\item $\safe_1\;(x := e) = ((\sec\;x = \hi) \vee \low\;e)$
\item $\fhigh\;(x := e) = (\sec\;x = \hi)$
\item $\safe_1\;(\Seq\;c_1\;c_2) = (\safe_1\;c_1 \wedge \safe_1\;c_2)$
\item $\fhigh\;(\Seq\;c_1\;c_2) = (\fhigh\;c_1 \wedge \fhigh\;c_2)$
\item $\safe_1\;(\If\;\tst\;c_1\;c_2) = 
\left\{
\begin{array}{ll}
   \safe_1\;c_1 \wedge \safe_1\;c_2, & \mbox{ if $\low\;\tst$}\\
   \safe_1\;c_1 \wedge \safe\;c_2 \wedge \fhigh\;c_1 \wedge \fhigh\;c_2, & \mbox{ otherwise}
\end{array}
\right.$
\item $\fhigh\;(\If\;\tst\;c_1\;c_2) = (\fhigh\;c_1 \wedge \fhigh\;c_2)$
\item $\safe_1\;(\While\;\tst\;c) = (\low\;\tst \wedge \safe_1\;c)$
\item $\fhigh\;(\While\;\tst\;c) = \False$
\item $\safe_1\;(\Par\;c_1\;c_2) = (\safe_1\;c_1 \wedge \safe_1\;c_2)$
\item $\low\;(\Par\;c_1\;c_2) = (\low\;c_1 \wedge \low\;c_2)$
\end{itemize}

Notice that the above clauses characterize the prediactes $\safe_1 : \com \ra \bool$ and $\fhigh : \com \ra \bool$ uniquely, 
i.e., could act as their definitions (recursively on the structure of commands).  
Since the predicate $\safe_1$ is stronger than $\fhigh$ (as its clauses are strictly stronger), we can remove $\safe_1\;c_1 \wedge \safe\;c_2$ from 
the ``otherwise" case of the $\If$ clause for $\safe_1$, obtaining:  
\begin{itemize}
\item $\safe_1\;(\If\;\tst\;c_1\;c_2) = 
\left\{
\begin{array}{ll}
   \safe_1\;c_1 \wedge \safe_1\;c_2, & \mbox{ if $\low\;\tst$}\\
   \fhigh\;c_1 \wedge \fhigh\;c_2, & \mbox{ otherwise}
\end{array}
\right. = 
\left\{
\begin{array}{ll}
   \safe_1\;c_1 \wedge \safe_1\;c_2, & \mbox{ if $\low\;\tst$}\\
   \fhigh\;(\If\;\tst\;c_1\;c_2), & \mbox{ otherwise}
\end{array}
\right.$
\end{itemize}

The clauses for $\safe_1$ and $\fhigh$ are now seen to coincide with our \cite[\S6]{cpp,jfr}  
clauses for $\ov{\wbist}$ and $\ov{\discr} \wedge \ov{\mayT}$, respectively, with the following variation: 
in \cite[\S6]{cpp,jfr} we do not commit to particular forms of  
tests or atomic statements, and therefore replace:  
\begin{itemize} 
\item $\low\;\tst$ with $\cpt\;\tst$
\item $\fhigh\;\atm$ with $\pres\;\atm$ (where $\atm$ is an atom, such as $x := e$)
\item $\safe_1\;\atm$ with $\cpt\;\atm$
\end{itemize}
Note that the predicates $\cpt$ and $\pres$, as defined in \cite[\S4]{cpp,jfr}, are semantic conditions expressed 
in terms of state indistinguishability, while $\low$, $\fhigh$ and $\safe_1$ are syntactic checks.  
than syntactic checks as here---the syntactic checks are easyly seen to be stronger, i.e., we have  
$\low\;\tst \LRA \cpt\;\tst$,  
$\fhigh\;\atm \LRA \pres\;\atm$ and 
$\safe_1\;\atm \LRA \cpt\;\atm$.  

\ \\
The main concurrent noninterference result from \cite{smiVol-multiThr}, Corollary 5.7, states (something slightly weaker than) 
the following: if $c ::_1 l$ for some $l \in L$, then $c \wbist c$. 
In the light of Lemma \ref{lem-tp1} and the above discussion, 
this result is subsumed by our Prop.~4 from \cite{cpp,jfr}, taking $\chi$ to be $\!\!\wbist\!\!$.  	

\ \\
For the rest of the type systems we discuss, we shall proceed with similar transformations at a higher pace.

\subsection{Volpano-Smith scheduler-independent noninterference} 

In \cite[\S7]{smiVol-multiThr}, another type system is defined, $::_2$, which has the same typing rules as $::_1$ except for the rule for $\If$, which 
is weakened by requiring the typing of the test to be $\lo$:\footnote{The same type system (except for the (PAR) rule) is introduced in \cite{volSmi-covert} for a sequential 
language with the purpose of preventing leaks through the covert channels of termination and exceptions.} 
\begin{center}
$\inferN{IF}
{\tst :: \lo \hspace*{3ex} c_1 ::_2 l \hspace*{3ex} c_2 ::_2 l}
{(\If\;\tst\;c_1\;c_2) ::_2 l}$
\end{center}

\begin{definition}\rm \label{def-safe2}
We define $\safe_2$ just like $\safe_1$, except for 
the case of $\If$, which becomes:
\begin{itemize}
\item $\safe_2\;(\If\;\tst\;c_1\;c_2) = ((\minTp\;\tst = \lo) \wedge \safe_2\;c_1 \wedge \safe_2\;c_2)$ 
\end{itemize}
\end{definition}

Similarly to Lemma \ref{lem-tp1}, we can prove:

\begin{lemma} \rm \label{lem-tp2}
The following are equivalent:
\\(1) $c ::_2 l$ 
\\(2) $\safe_2\;c$ and $l \leq \maxTp_1\;c$.
\end{lemma}

The inferred clauses for $\safe_2$ are the same as those for $\safe_1$, except for 
the one for $\If$, which becomes: 
\begin{itemize}
\item $\safe_2\;(\If\;\tst\;c_1\;c_2) = (\low\;\tst \wedge \safe_2\;c_1 \wedge \safe_2\;c_2)$ 
\end{itemize}

Then $\safe_2$ is seen to coincide with $\ov{\siso}$ from \cite[\S6]{cpp,jfr}.  

\ \\
In \cite{smiVol-multiThr} it is proved (via Theorem 7.1) that the soundness result for $::_1$ also holds for $::_2$.  
In fact, one can see that Theorem 7.1 can be used to prove something much 
stronger: if $c ::_2 l$ for some $l \in L$, then $\siso\;c$.  
This result is subsumed by our Prop.~4 from \cite{cpp,jfr}, taking $\chi$ to be $\siso$.

\subsection{Boudol-Castellani termination-insensitive noninterference}

As we already discussed in \cite{cpp,jfr}, 
Boudol and Castellani \cite{boudCast0,boudCast} work on improving the harsh Vopano-Smith typing of $\While$ 
(which requires low tests), but they pay a (comparatively small) price in terms of typing sequential composition, where 
what the first command reads is required to be below what the second command writes.  (Essentially the same type system is introduced independently 
by Smith \cite{smi-NewTypeSys,smi-weakProbBis} for studying probabilistic noninterference in the presence of uniform scheduling.  
Boudol and Castellani, as well as Smith, consider parallel composition only at the top level. Barthe and Nieto \cite{barthe-isabelle} 
raise this restriction, allowing nesting $\Par$ inside other language constructs, as we do here.)

To achieve this, they type commands $c$ to a pair of security levels $(l,l')$: the contravariant ``write" type $l$ 
(similar to the Volpano-Smith one) and an extra covariant ``read" type $l'$.         

\begin{center}
$\inferN{ASSIGN}
{\sec\;x = l \hspace*{3ex} e :: l}
{(x := e) ::_2 (l,l')}$
\hspace*{5ex}
$\inferN{COMPOSE}
{c_1 ::_3 (l_1,l_1') \hspace*{3ex} c_2 ::_3 (l_2,l_2')  \hspace*{3ex} l_1' \leq l_2}
{(\Seq\;c_1\;c_2) ::_3 (l_1 \wedge l_2, l_1' \vee l_2')}$
\\
\vspace*{1.3ex} 
$\inferN{IF}
{\tst :: l_0 \hspace*{2.7ex} c_1 ::_3 (l,l') \hspace*{2.7ex} c_2 ::_3 (l,l')  \hspace*{2.7ex} l_0 \leq l}
{(\If\;\tst\;c_1\;c_2) ::_3 (l, l_0 \vee l')}$
\hspace*{1ex} 
$\inferN{WHILE}
{\tst :: l' \hspace*{2.7ex} c ::_3 (l,l') \hspace*{2.7ex} l' \leq l}
{(\While\;\tst\;c) ::_3 (l,l')}$
\\\vspace*{1.3ex}
$\inferN{PAR}
{c_1 ::_3 l \hspace*{3ex} c_2 ::_3 l}
{(\Par\;c_1\;c_2) ::_3 l}$
\hspace*{15ex}  
$\inferN{SUBTYPE}
{c ::_3 (l_1,l_1') \hspace*{3ex} l_2 \leq l_1 \hspace*{3ex} l_1' \leq l_2'}
{c ::_3 (l_2,l_2')}$
\end{center}

We think of $c ::_3 (l,l')$ as saying: 
\begin{itemize}
\item There is no downwards flow in $c$. 
\item $l$ is a lower bound on the level of the variables that the execution of $c$ writes to. 
\item $l'$ is an upper bound on the level of the variables that $c$ reads, more precisely, 
that the control flow of the execution of $c$ depends on.   
\end{itemize}
(This intuition is accurately reflected by Lemma \ref{lem-tp3} below.)

In \cite{boudCast0,boudCast}, the rule for $\While$ is slightly different, namely: 
\begin{center}
$\inferN{WHILE'}
{\tst :: l_0 \hspace*{3ex} c ::_3 (l,l') \hspace*{3ex} l_0 \vee l' \leq l}
{(\While\;\tst\;c) ::_3 (l,l_0 \vee l')}$
\end{center}
However, due to subtyping, it is easily seen to be equivalent to the one we listed. 
Indeed: 
\begin{itemize}
\item (WHILE) is an instance of (WHILE') taking $l_0 = l'$. 
\item Conversely, (WHILE') follows from (WHILE) as follows: Assume the hypotheses of (WHILE').  
By subtyping, we have $\tst :: l_0 \vee l'$ and $c ::_3 (l,l_0 \vee l')$, 
hence, by (WHILE), we have $(\While\;\tst\;c) ::_3 (l,l_0 \vee l')$, as desired.  
\end{itemize}

Following for $::_3$ the same technique as in the case of $::_1$ and $::_2$, we define the 
functions $\maxWtp : \com \ra L$ (read ``maximum writing type") 
and $\minRtp : \com \ra L$ (read ``minimum reading type")
and the predicate $\safe_3 : \com \ra \bool$:  

\begin{definition}  \rm\label{def-safe3}
\begin{itemize}
\item $\safe_3\;(x := e) = (\minTp\;e \leq \sec\;x)$
\item $\maxWtp\;(x := e) = \sec\;x$ 
\item $\minRtp\;(x := e) = \lo$
\item $\safe_3\;(\Seq\;c_1\;c_2) = (\safe_3\;c_1 \wedge \safe_3\;c_2 \wedge (\minRtp\;c_1 \leq \maxWtp\;c_2))$
\item $\maxWtp\;(\Seq\;c_1\;c_2) = (\maxWtp\;c_1 \wedge \maxWtp\;c_2)$
\item $\minRtp\;(\Seq\;c_1\;c_2) = (\minRtp\;c_1 \vee \minRtp\;c_2)$
\item $\safe_3\;(\If\;\tst\;c_1\;c_2) = (\safe_3\;c_1 \wedge \safe_3\;c_2 \wedge (\minTp\;\tst \leq (\maxWtp\;c_1 \wedge \maxWtp\;c_2)))$
\item $\maxWtp\;(\If\;\tst\;c_1\;c_2) = (\maxWtp\;c_1 \wedge \maxWtp\;c_2)$
\item $\minRtp\;(\If\;\tst\;c_1\;c_2) = (\minTp\;\tst \vee \minRtp\;c_1 \vee \minRtp\;c_2)$
\item $\safe_3\;(\While\;\tst\;c) = (\safe_3\;c \wedge ((\minTp\;\tst \vee \minRtp\;c) \leq \maxWtp\;c))$
\item $\maxWtp\;(\While\;\tst\;c) = \maxWtp\;c$
\item $\minRtp\;(\While\;\tst\;c) = (\minTp\;\tst \vee \minRtp\;c)$
\item $\safe_3\;(\Par\;c_1\;c_2) = (\safe_3\;c_1 \wedge \safe_3\;c_2)$
\item $\maxWtp\;(\Par\;c_1\;c_2) = (\maxWtp\;c_1 \wedge \maxWtp\;c_2)$
\item $\minRtp\;(\Par\;c_1\;c_2) = (\minRtp\;c_1 \vee \minRtp\;c_2)$
\end{itemize}
\end{definition}

Furthermore, similarly to the cases of $\safe_1$ and $\safe_2$, we have that: 

\begin{lemma} \rm \label{lem-tp3}
The following are equivalent:
\\(1) $c ::_3 (l,l')$ 
\\(2) $\safe_3\;c$ and $l \leq \maxWtp\;c$ and $\minRtp\;c \leq l'$.
\end{lemma}

\ \\
Now, let us write:
\begin{itemize}
\item $\high\;c$, for the sentence $\maxWtp\;c = \hi$
\item $\low\;c$, for the sentence $\minRtp\;c = \lo$
\end{itemize}

Then, immediately from the definitions of $\maxWtp$ and $\minRtp$, we have the following:

\begin{itemize}
\item $\safe_3\;(x := e) = ((\sec\;x = \hi) \vee \low\;e)$ 
\item $\high\;(x := e) = (\sec\;x = \hi)$ 
\item $\low\;(x := e) = \True$
\item $\safe_3\;(\Seq\;c_1\;c_2) = (\safe_3\;c_1 \wedge \safe_3\;c_2 \wedge (\low\;c_1 \vee \high\;c_2))$
\item $\high\;(\Seq\;c_1\;c_2) = (\high\;c_1 \wedge \high\;c_2)$
\item $\low\;(\Seq\;c_1\;c_2) = (\low\;c_1 \wedge \low\;c_2)$
\item $\safe_3\;(\If\;\tst\;c_1\;c_2) = (\safe_3\;c_1 \wedge \safe_3\;c_2 \wedge (\low\;\tst \vee (\high\;c_1 \wedge \high\;c_2)))$
\item $\high\;(\If\;\tst\;c_1\;c_2) = (\high\;c_1 \wedge \high\;c_2)$
\item $\low\;(\If\;\tst\;c_1\;c_2) = (\low\;\tst \wedge \low\;c_1 \wedge \low\;c_2)$
\item $\safe_3\;(\While\;\tst\;c) = (\safe_3\;c \wedge ((\low\;\tst \wedge \low\;c) \vee \high\;c))$
\item $\high\;(\While\;\tst\;c) = \high\;c$
\item $\low\;(\While\;\tst\;c) = (\low\;\tst \wedge \low\;c)$
\item $\safe_3\;(\Par\;c_1\;c_2) = (\safe_3\;c_1 \wedge \safe_3\;c_2)$
\item $\high\;(\Par\;c_1\;c_2) = (\high\;c_1 \wedge \high\;c_2)$
\item $\low\;(\Par\;c_1\;c_2) = (\low\;c_1 \wedge \low\;c_2)$ 
\end{itemize}

Then $\high$ and $\low$ are stronger than $\safe_3$, and hence we can rewrite the $\Seq$, $\If$ and $\While$ clauses 
for $\safe_3$ as follows:
\begin{itemize}
\item $\safe_3\;(\Seq\;c_1\;c_2) = ((\low\;c_1 \wedge \safe_3\;c_2) \vee (\safe_3\;c_1 \wedge \high\;c_2))$
\item $\safe_3\;(\If\;\tst\;c_1\;c_2) = 
\left\{
\begin{array}{ll}
   \safe_3\;c_1 \wedge \safe_3\;c_2, & \mbox{ if $\low\;\tst$}\\
   \high\;c_1 \wedge \high\;c_2, & \mbox{ otherwise}
\end{array}
\right. = 
\left\{
\begin{array}{ll}
   \safe_3\;c_1 \wedge \safe_3\;c_2, & \mbox{ if $\low\;\tst$}\\
   \high\;(\If\;\tst\;c_1\;c_2), & \mbox{ otherwise}
\end{array}
\right.$
\item $\safe_3\;(\While\;\tst\;c) = ((\low\;\tst \wedge \low\;c) \vee \high\;c) = (\low\;(\While\;\tst\;c) \vee \high\;(\While\;\tst\;c))$
\end{itemize}

The clauses for $\safe_3$, $\high$ and $\low$ are now seen to coincide with our \cite[\S6]{cpp,jfr}  
clauses for $\ov{\zobis}$ and $\ov{\discr}$ and $\ov{\siso}$, respectively.  

\ \\
The main concurrent noninterference result from \cite{boudCast0,boudCast}  
(Theorem 3.13 in \cite{boudCast0} and Theorem 3.16 in \cite{boudCast}), states  
(something slightly weaker than) the following: if $c ::_3 l$ for some $l \in L$, then $c \zobis c$. 
In the light of Lemma \ref{lem-tp3} and the above discussion, 
this result is subsumed by our Prop.~4 from \cite{cpp,jfr}, taking $\chi$ to be $\!\!\zobis\!\!$.

\subsection{Matos and Boudol's further improvement}

Mantos and Boudol \cite{matos-OnDeclass0,matos-OnDeclass,boudImproved} study a richer language than the one we consider here, namely, 
an ML-like language. Moreover, they also consider a declassification construct.  
We shall ignore these extra features and focus on the restriction of their results to our simple while language. 
Moreover, they parameterize their development by a set of strongly terminating expressions (commands in our setting)---here we 
fix this set to be that of commands not containing while loops.  

The type system $::_4$ from \cite{matos-OnDeclass0,matos-OnDeclass,boudImproved} is based on a refinement of $::_3$, noticing that, 
as far as the reading type goes,  
one does not care about all variables a command reads (i.e., the variables that affect the control flow of its execution), 
but can restrict attention to those that may affect the {\em termination} of its execution.  

The typing rules of $::_4$ are identical to those of $::_3$, except for the $\If$ rule, which becomes: 

\begin{center}
$\inferN{IF}
{\tst :: l_0 \hspace*{2.7ex} c_1 ::_3 (l,l') \hspace*{2.7ex} c_2 ::_3 (l,l')  \hspace*{2.7ex} l_0 \leq l}
{(\If\;\tst\;c_1\;c_2) ::_3 (l, k)}$
\end{center}
where 
$k = 
\left\{
\begin{array}{ll}
   \lo, & \mbox{ if $c_1,c_2$ do not contain $\While$ subexpressions}\\
   l_0 \vee l', & \mbox{ otherwise}
\end{array}
\right.$

We think of $c ::_4 (l,l')$ as saying: 
\begin{itemize}
\item There is no downwards flow in $c$. 
\item $l$ is a lower bound on the level of the variables that the execution of $c$ writes to. 
\item $l'$ is an upper bound on the level of the variables that $c$ termination-reads, i.e., 
that termination of the execution of $c$ depends on.   
\end{itemize}
%

(In \cite{matos-OnDeclass0,matos-OnDeclass,boudImproved}, 
$\While$ is not a primitive, but is derived from higher-order recursion---however, the effect of the higher-order typing system on $\While$ 
is the same as that of our $::_3$, as shown in \cite{matos-OnDeclass}.  Moreover, due to working in a functional language with side effects, 
\cite{matos-OnDeclass0,matos-OnDeclass,boudImproved} record not two, but three security types: in addition to our $l$ and $l'$ (called there 
the writing and termination effects, respectively), they also record $l''$ (called there the reading effect) 
which represents an upper bound on the security levels of variables the returned value of $c$ depends on---here, this information is unnecessary, 
since $c$ returns no value.)

\begin{definition}  \rm\label{def-safe4}
We define the 
function $\minTRtp : \com \ra L$ (read ``minimum termination-reading type")
and the predicate $\safe_4 : \com \ra \bool$ as follows: 
$\minTRtp$ is defined using the same recursive clauses as $\minRtp$, except for the clause for $\If$, which becomes: 
\begin{itemize}
\item $\minTRtp\;(\If\;\tst\;c_1\;c_2) =$
\\ 
$\left\{
\begin{array}{ll}
   \lo, & \mbox{ if $c_1,c_2$ do not contain $\While$ subexpressions}\\
   \minTp\;\tst \vee \minTRtp\;c_1 \vee \minTRtp\;c_2, & \mbox{ otherwise}
\end{array}
\right.
$
\end{itemize}
$\safe_4$ is defined using the same clauses as $\safe_3$ with $\minTRtp$ replacing $\minRtp$.  
\end{definition} 

\begin{lemma} \rm \label{lem-tp4}
The following are equivalent:
\\(1) $c ::_4 (l,l')$ 
\\(2) $\safe_4\;c$ and $l \leq \maxWtp\;c$ and $\minTRtp\;c \leq l'$.
\end{lemma}

\ \\
Now, let us write:
\begin{itemize}
\item $\wlow\;c$ (read ``$c$ has low tests on top of while subexpressions"), for the sentence $\minTRtp\;c = \lo$
\item $\noWhile\;c$, for the sentence ``$c$ contains no $\While$ subexpressions"
\end{itemize}

We obtain: 

\begin{itemize}
\item $\safe_4\;(x := e) = ((\sec\;x = \hi) \vee \low\;e)$  
\item $\wlow\;(x := e) = \True$
\item $\safe_4\;(\Seq\;c_1\;c_2) = (\safe_4\;c_1 \wedge \safe_4\;c_2 \wedge (\wlow\;c_1 \vee \high\;c_2))$
\item $\wlow\;(\Seq\;c_1\;c_2) = (\wlow\;c_1 \wedge \wlow\;c_2)$
\item $\safe_4\;(\If\;\tst\;c_1\;c_2) = (\safe_4\;c_1 \wedge \safe_4\;c_2 \wedge (\wlow\;\tst \vee (\high\;c_1 \wedge \high\;c_2)))$
\item $\wlow\;(\If\;\tst\;c_1\;c_2) = (\low\;\tst \wedge \wlow\;c_1 \wedge \wlow\;c_2) \vee (\noWhile\;c_1 \wedge \noWhile\;c_2)$
\item $\safe_4\;(\While\;\tst\;c) = (\safe_4\;c \wedge ((\low\;\tst \wedge \low\;c) \vee \high\;c))$
\item $\wlow\;(\While\;\tst\;c) = (\low\;\tst \wedge \wlow\;c)$
\item $\safe_4\;(\Par\;c_1\;c_2) = (\safe_4\;c_1 \wedge \safe_4\;c_2)$
\item $\wlow\;(\Par\;c_1\;c_2) = (\wlow\;c_1 \wedge \wlow\;c_2)$ 
\end{itemize}

We can prove by induction on $c$ that $\safe_1\;c = (\safe_4\;c \wedge \wlow\;c)$
Using this, we rewrite the $\Seq$, $\If$ and $\While$ clauses 
for $\safe_4$ as follows:
\begin{itemize}
\item $\safe_4\;(\Seq\;c_1\;c_2) = ((\safe_1\;c_1 \wedge \safe_4\;c_2) \vee (\safe_4\;c_1 \wedge \high\;c_2))$
\item $\safe_4\;(\If\;\tst\;c_1\;c_2) = 
\left\{
\begin{array}{ll}
   \safe_4\;c_1 \wedge \safe_4\;c_2, & \mbox{ if $\low\;\tst$}\\
   \high\;(\If\;\tst\;c_1\;c_2), & \mbox{ otherwise}
\end{array}
\right.$
\item $\safe_4\;(\While\;\tst\;c) = (\safe_1\;(\While\;\tst\;c) \vee \high\;(\While\;\tst\;c))$
\end{itemize} 

Then $\safe_4$ turns out to coincide with our $\ov{\wbis}$ from \cite[\S6]{cpp,jfr}.  

\ \\
The main noninterference result from \cite{matos-OnDeclass0,matos-OnDeclass,boudImproved}  
(in \cite{boudImproved}, the soundness theorem in \S5), states the following: if $c ::_4 l$ for some $l \in L$, then $c \wbis c$. 
In the light of Lemma \ref{lem-tp3} and the above discussion, 
this result is subsumed by our Prop.~4 from \cite{cpp,jfr}, taking $\chi$ to be $\!\!\wbis\!\!$.




\bibliography{tex,mantelsudbrock}

\end{document}